\documentclass[apjl]{emulateapj}

\newcommand{\myemail}{reggiani@phys.ethz.ch}

\slugcomment{ }

\shorttitle{Discovery of a companion candidate in the HD169142 transition disk}
\shortauthors{Reggiani et al.}

\begin{document}

\title{Discovery of a companion candidate in the HD169142 transition disk \\ and the possibility of multiple planet formation}

\author{Maddalena Reggiani\altaffilmark{1}}
\author{Sascha P. Quanz\altaffilmark{1}}
\author{Michael R. Meyer\altaffilmark{1}}
\author{Laurent Pueyo\altaffilmark{2}}
\author{Olivier Absil\altaffilmark{3}}
\author{Adam Amara\altaffilmark{1}}
\author{Guillem Anglada\altaffilmark{4}}
\author{Henning Avenhaus\altaffilmark{1}}
\author{Julien H. Girard\altaffilmark{5}}
\author{Carlos Carrasco Gonzalez\altaffilmark{6}}
\author{Graham James\altaffilmark{7}}
%\author{Bruce M. Macintosh\altaffilmark{8}}
\author{Dimitri Mawet\altaffilmark{5}}
\author{Farzana Meru\altaffilmark{1}}
\author{Julien Milli\altaffilmark{5}}
\author{Mayra Osorio\altaffilmark{4}}
\author{Schuyler Wolff\altaffilmark{2}}
\author{Jose-Maria Torrelles\altaffilmark{8}}

\altaffiltext{1}{Institute for Astronomy, ETH Zurich, CH-8093 Zurich, Switzerland; \myemail}
\altaffiltext{2}{Space Telescope Science Institute, 3700 San Martin Dr, Baltimore, MD 21218, United States}
\altaffiltext{3}{D\'epartement d'Astrophysique, G\'eophysique et Oc\'eanographie, Universit\'e de Li\`ege, 17 All\'ee du Six Ao\^ut, 4000 Li\`ege, Belgium}
\altaffiltext{4}{Instituto de Astrof\'isica de Andaluc\'ia, CSIC, Glorieta de la Astronom\'{\i}a s/n, 18008 Granada, Spain}
\altaffiltext{5}{European Southern Observatory, Alonso de Cordova 3107, Casilla 19001, Vitacura, Santiago 19, Chile}
\altaffiltext{6}{Centro de Radioastronom\'{\i}a y Astrof\'{\i}sica (UNAM), Apartado Postal 3-72 (Xangari), 58089 Morelia, Mexico}
\altaffiltext{7}{University of California, 644 Campbell Hall, Berkeley}
%\altaffiltext{8}{Kavli Institute for Particle Astrophysics and Cosmology, Stanford University, Stanford, CA 94305} 
\altaffiltext{8}{Instituto de Ciencias del Espacio (CSIC)-UB/IEEC, Universitat de Barcelona, Mart\'i i Franqu\`es 1, 08028 Barcelona, Spain}

\begin{abstract}
We present $L'$ and $J$-band high-contrast observations of HD169142, obtained with the VLT/NACO AGPM vector vortex coronagraph and the Gemini Planet Imager, respectively. A source located at 0''.156$\pm$0''.032 north of the host star ($PA$=7.4$^{\circ}\pm$11.3$^{\circ}$) appears in the final reduced $L'$ image. At the distance of the star ($\sim$145 pc), this angular separation corresponds to a physical separation of 22.7$\pm$4.7 AU, locating the source within the recently resolved inner cavity of the transition disk. The source has a brightness of $L'$=12.2$\pm$0.5 mag, whereas it is not detected in the $J$ band ($J>$13.8 mag). If its $L'$ brightness arose solely from the photosphere of a companion and given the $J-L'$ color constraints, it would correspond to a 28-32 $M_{Jupiter}$ object at the age of the star, according to the COND models. Ongoing accretion activity of the star suggests, however, that gas is left in the inner disk cavity from which the companion could also be accreting. In this case the object could be lower in mass and its luminosity enhanced by the accretion process and by a circumplanetary disk. A lower mass object is more consistent with the observed cavity width. Finally, the observations enable us to place an upper limit on the $L'$-band flux of a second companion candidate orbiting in the disk annular gap at $\sim$50 AU, as suggested by millimeter observations. If the second companion is also confirmed, HD169142 might be forming a planetary system, with at least two companions opening gaps and possibly interacting with each other.

\end{abstract}


\section{Introduction}
To understand how planet formation proceeds and in which chemical and physical conditions it occurs, young, gas rich disks have been studied by numerous observing programs. The so-called ``transition disks" show inner holes, bright rims and annular gaps, which could be tracing on-going planet formation \citep{Espaillat2014}.

A few planet candidates have been found in these disks. Some of them were detected with sparse aperture masking (SAM) observations in the disk gaps around their host stars \citep[e.g. LkCa15 b, TCha b;][]{Kraus2012,Huelamo2011}, but disk features or scattered light from inner disk rims have been suggested to explain some of these observations \citep[e.g.][]{Cieza2013,Thalmann2014}. Coronagraphy supported angular differential imaging \citep[ADI,][]{Marois2006} and spectroastrometry of CO rovibrational lines provided direct empirical evidence for two companions orbiting the young star HD100546 \citep{Quanz2013a,Brittain2013}.

In this letter we present $L'$- and $J$-band high-contrast imaging observations of HD169142\footnote{Based on observations made with ESO Telescopes at the Paranal Observatory under program 291.C-5020(A) and with Gemini during GPI Early Science runs under program GS-2014A-SV-407}, which reveal the presence of a companion candidate in the inner cavity of its transition disk.

HD169142 is a young Herbig Ae/Be star (see Table~\ref{table:1} for stellar properties) with a complex circumstellar disk structure, that has been extensively studied \citep[e.g.,][]{Dent2006,Grady2007,Meeus2010,Honda2012,Quanz2013b}. 
It has a small central disk ($<$0.7 AU) with either a hot-dust halo \citep{Honda2012} or a hot inner wall \citep{Osorio2014}, an inner cavity, a bright rim ($\sim$25 AU), and an annular gap from 40 to 70 AU. The latter features have recently been resolved with polarimetric differential imaging (PDI) in the H band \citep{Quanz2013b}.
Furthermore, EVLA 7-mm observations of the disk detected the thermal dust emission of the bright rim, seen in polarized light, and revealed a compact emission source inside the annular gap \citep[at $\sim$50 AU;][]{Osorio2014}. These recent observations as well as the morphology of the disk suggest that HD169142 could be hosting young planetary companions.

\begin{table}
\begin{center}
\caption{Stellar Parameters.\label{table:1}}
\begin{tabular}{crc}
\tableline\tableline
Parameter & Value & Reference \\ [0.5ex]
\tableline
    R.A. (J2000) & 18h 24m 29.785s &   [1]\\
    Dec. (J2000) & -29$^\circ$ 46' 49.829''  &   [1]\\
    Distance (pc) & 145-151  &   [6]; [3]\\
    J (mag) & 7.31$\pm$0.02 &   [1]\\ 
    H (mag) & 6.91$\pm$0.04 &   [1]\\ 
    K$_s$ (mag) &  6.41$\pm$0.02  &  [1]\\ 
    L' (mag) &  5.66$\pm$0.03  &   [7]\\ 
    Sp. Type &  A9III/IVe/A7V &   [2]; [3]\\
     $v$ $sin$ $i$ (km s$^{-1}$) & 55$\pm$5 &   [2]\\
      Age (Myr) & 1-5/12/3-12 & [2]; [3]; [4] \\
     T$_{eff}$ & 7500$\pm$200/6500/7650$\pm$150 &  [2]; [3]\\
      Mass ($M_{\odot}$) & $\sim$1.65 & [3]  \\        
       $L_{*}$ ($L_{\odot}$)& $\sim$8.6-13 &  [3]; [8]\\
       $R_{*}$ ($R_{\odot}$) & $\sim$1.6 & [3]; [5] \\
       $\dot{M}$ (10$^{-9}M_{\odot}  yr^{-1}$) & $\sim$3.1/$\leq 1.25\pm 0.55$ & [3]; [4] \\
       $\log$ $g$  & 3.7 $\pm$0.1/4.0-4.1 & [2]; [5] \\ \hline
\tableline
        \end{tabular}

{\footnotesize $References$: [1] From 2MASS point source catalog \citep{Cutri2003} and corrected for proper motions to the epoch of our VLT observations; [2] \cite{Guimaraes2006}; [3] \cite{Blondel2006}; [4] \cite{Grady2007}; [5] \cite{Meeus2010}; [6] \cite{Sylvester1996}; [7] \cite{vanderVeen1989}; [8] \cite{Marinas2011} }
\end{center}
\end{table}

\section{Observations and Data Reduction}
\subsection{NACO L' band}
HD169142 was observed on June 28, 2013 with the VLT/NACO annular groove phase mask (AGPM) vector vortex coronagraph \citep{Mawet2013} in pupil stabilized mode. All images were taken with the L27 camera (plate scale $\sim$27.15 mas pixel$^{-1}$) using the $L'$ filter ($\lambda_c=3.8\mu m$, $\Delta\lambda=0.62\mu m$).
The detector reads were recorded in ``cube" mode and the integration time per read was set to 0.25 s. We obtained 111 minutes of on-source integration time and $\sim$159$^{\circ}$ of field rotation. The sky was observed every $\sim20$ minutes and unsaturated images (0.05 s) of the star were acquired to calibrate the photometry. 
Table~\ref{table:2} summarizes the observations.  

\begin{table*}
\begin{center}
\caption{Summary of Observations.\label{table:2}}
\begin{tabular}{cccccc}
\tableline\tableline
Instrument & Filter & No. of detector reads $\times$ exp. time & No. of data cubes & Parallactic angle start/end & Airmass range \\ 
\tableline
    VLT/NACO & L' & 60 $\times$ 0.25$s$ & 444 & -84.29/74.70 & 1.097 -1.038\\
    Gemini/GPI & J-coro & 1 $\times$ 60$s$ & 52 & -96.65/-102.96 & 1.048 -1.001\\ \hline
\tableline
        \end{tabular}
\end{center}
\end{table*}

We subtracted the background from each frame adopting for each cube the mean of the closest sky measures in time.
 We applied a bad pixel/cosmic ray correction, adopting a 5-sigma threshold and replacing every anomalous pixel with the mean value of the 8 surrounding pixels. Since the AGPM already requires a mandatory centering accuracy of $\sim$0.3 pixel, which has been checked every 10-30 minutes during the observations, we did not apply any further centering to our images. As a test, additional centering was performed in a second data reduction pipeline (see Section~\ref{Results}), yielding consistent results. Finally from each image we created a $\sim$2"$\times$2'' sub-image (75$\times$75 pixels) centered on the star, resulting in a stack of $\sim$25000 sub-images.

To subtract the stellar PSF from all the sub-images we used the principal component analysis (PCA) based package \texttt{PYNPOINT}\footnote{Public version of the code can be found at \url{http://pynpoint.ethz.ch} and additional material in Amara, Quanz, and Akeret, Astronomy and Computing (submitted).} \citep{Amara2012}. 
\texttt{PYNPOINT} creates a set of orthogonal basis functions to reproduce the stellar PSF and fits it to the individual frames with a chosen number of PCA coefficients. It subtracts the PSF from each frame, de-rotates the frames to the same field rotation, averages them and convolves them with a Gaussian kernel (0.5$\times$FWHM$_{\rm PSF}$) to get the final image of the stack. We adopted the 20 PCA \texttt{PYNPOINT} image as final reference, as it shows lower residual noise compared to images obtained with higher or lower numbers of PCA coefficients.

\subsection{GPI J band}
Spectral data were obtained in $J$ band (1.12-1.35 $\mu m$, $R=$35-39) on April 26, 2014 with the Gemini Planet Imager \citep[GPI,][]{Macintosh2014} using an apodized Lyot coronagraph in pupil stabilized mode (see Table~\ref{table:2}). The total on source integration time was 52 minutes with a total field rotation of $\sim$6$^{\circ}$.

We adopted the GPI pipeline \citep{Maire2010,Maire2012} for bad-pixel removal, de-striping, flat-fielding, wavelength calibration, and to convert the data into spectral data cubes. In total, the data are made of 52 cubes, consisting of 37 spectral channels each. The cubes were then aligned and registered according to the procedure described in \cite{Crepp2011} and \cite{Pueyo2012} in order for the data to be placed in a frame where the speckles scale is fixed as a function of wavelength while putative point sources move radially. The KLIP algorithm was then carried out as presented in \cite{Soummer2012}, using the variations for IFS data described in Pueyo et al 2014 (in review). Finally, we reduced each slice of each cube separately and then mean combined them both in wavelength and time.

\begin{figure*}
\includegraphics[angle=0,scale=.36]{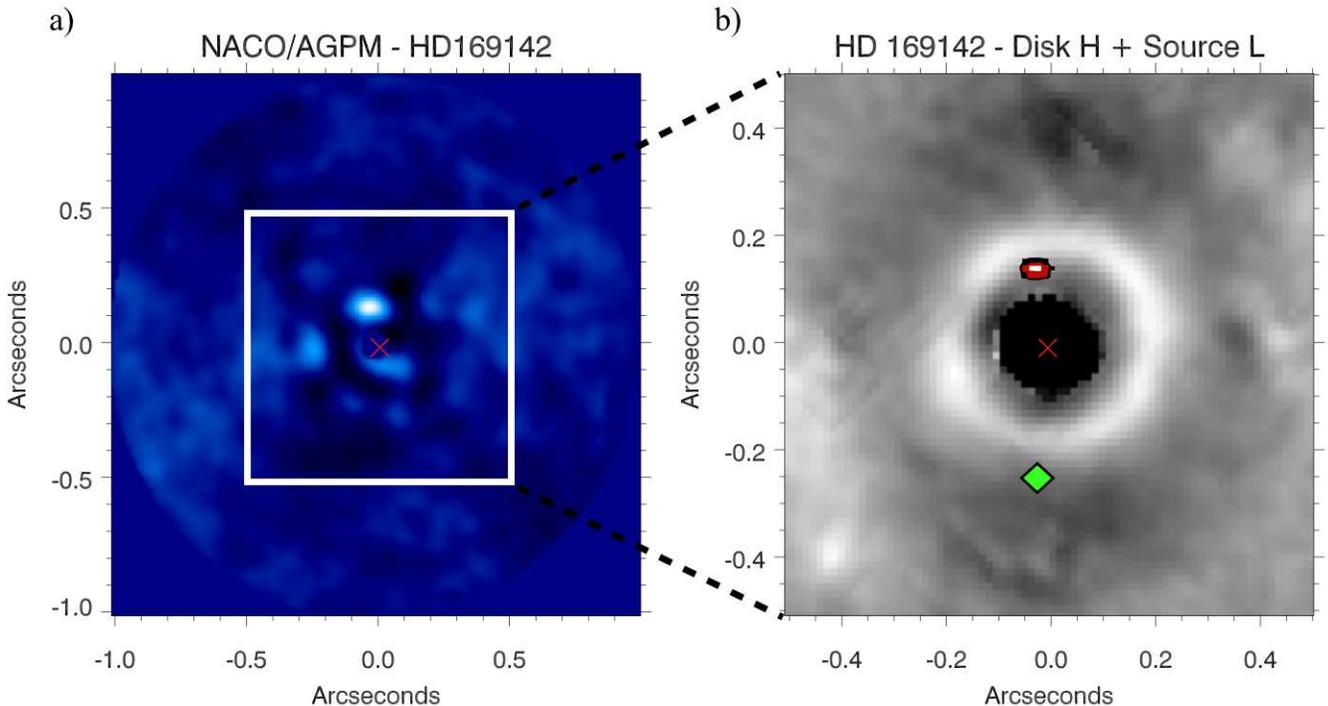}
\caption{a) NACO/AGPM $L'$ image of HD169142, using \texttt{PYNPOINT} with 20 PCA coefficients. A bright source is detected north of the central star. The image is scaled with respect to the maximum flux. b) $H$-band PDI image of the circumstellar disk of HD169142 \citep{Quanz2013b}. The inner cavity ($<$25 AU), the bright rim, and the annular gap (40-70 AU) are clearly visible. Overplotted in red contours is the detected $L'$ source. The green diamond indicates the location of the compact 7-mm emission detected by \cite{Osorio2014}.\label{figure:1}}
\end{figure*}

\section{Results}\label{Results}
\subsection{Detection of an Emission Source in the L' band}
In the final $L'$-band image an emission source is revealed north of HD169142 (see Figure~\ref{figure:1}a). To assess its reliability, we performed a series of tests.
\begin{enumerate}
\item We varied the number of PCA coefficients used in \texttt{PYNPOINT} between 5 and 120.
\item We divided the dataset into different subsets containing either half or a third of the frames, but spanning the full field rotation.
\item We did two ``blind" data reductions to confirm the result using both a separate PCA-based pipeline \citep{Absil2013,Mawet2013} and the LOCI algorithm \citep{Lafreniere2007}.
\end{enumerate}
In each case, we always found a bright emission source at the same location. 
Since the residual speckle noise does not follow a Gaussian distribution, the calculation of Gaussian confidence levels may not be appropriate \citep[see e.g.][]{Kasper2007}.
To estimate the statistical confidence of the detection we used the final image and selected 28 pixels in 2 concentric rings around the star as noise reference. 10 pixels had the same separation from the star as the peak flux of the companion, the ring of the other 18 pixels had a radius of 0''.23 and included a bright residual feature east of the central star. These pixels were chosen to be statistically independent in the convolved image and were used to compute the mean, variance and skewness of the distribution and built a probability density function (PDF) assuming a log-normal underlying distribution. From this PDF we estimated the likelihood of finding a pixel value equaling the companion's peak flux or higher to be  $p<0.2 \%$. 

The results of all these tests give us confidence that the detection is real. None of the other features in the final image is a reliable detection, based on these tests.

To derive the astrometry and photometry of the source, we inserted negative artificial planets in the individual exposures varying at the same time their brightness (with steps of 0.25 mag) and location (with steps of 0.25 pixel) and then re-ran \texttt{PYNPOINT}. To generate them we used a MonteCarlo photon generator with customizable FWHM. We adopted the FWHM measured from the unsaturated images of the photometric calibration dataset and we scaled the flux of the objects relative to the star, accounting for the difference in exposure time. We performed some simulations to evaluate the astrometric distortion induced by the vortex. At 1 $\lambda/D$, the offset is 0.075 FWHM, well below the speckle noise induced errors, and thus negligible. Each time, we used the final \texttt{PYNPOINT} image to calculate the deviation of the remaining flux at the object's location compared to the background noise in an annulus of 1 FWHM around the detection. We chose as brightness and astrometry of the source the combination of flux and position that yields the lowest deviation; i.e. the best subtraction. 

\begin{figure}
\includegraphics[angle=0,scale=.36]{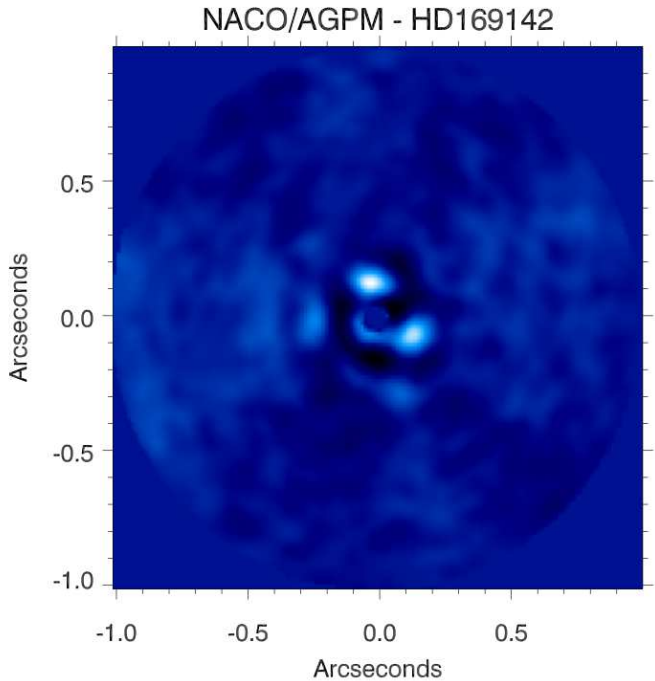}
\caption{NACO/AGPM $L'$ image of HD169142 with an artificial planet of the same brightness and angular separation as the detection. This image shows the final outcome of \texttt{PYNPOINT} with 20 PCA coefficients. Besides the bright source detected north of the central star, we could recover the artificial planet at $PA\simeq$ 270$^{\circ}$.\label{figure:3}}
\end{figure}

To conclude, the source is located at 0".156$\pm$0''.032 from the central star at a position angle of $PA$=7.4$^{\circ}\pm$11.3$^{\circ}$. Our best estimate of the contrast for the object is $\Delta L'$=6.5$\pm$0.5 mag. The errors on these measurements are the 1-$\sigma$ deviation quantities.  These estimates are consistent with the expected performance of the AGPM at $\sim$0''.16 \citep[see e.g.][]{Mawet2013}. We also inserted an artificial positive planet of the same brightness at the same separation but at a different position angle and we were able to recover it (see Figure~\ref{figure:3}). The  artificial planet appears elongated, showing that the final shape of point-like sources at these small separations is affected by image processing. 
The observed magnitude for HD169142 is $L'$=5.66$\pm$0.03 mag \citep{vanderVeen1989}. Thus, we derived an apparent magnitude of  $L'$=12.2$\pm$0.5 mag for the newly detected source. The uncertainty is the square-root of the sum of squares of the errors on the stellar and the object's magnitudes.

An independent confirmation of this detection with the same instrument and at the same wavelength is provided by Biller et al., submitted. The astrometry and photometry of the detections are consistent within the errors.

\subsection{Non-Detection in the $J$ band}
No source was detected in the $J$ band images. We derived the 5-$\sigma$ detection limits by match filtering the reduced data using empirical PSF templates based on the astrometric spots. We then set as a conservative upper limit the maximum of the matched filtered PSF in various annuli around the star and corrected for self subtraction.  Given the small separation of the $L'$ band candidate, a trade between self-subtraction and speckle suppression was found, based on synthetic companions injected in the dataset at 6 azimuthal zones of radius 20 pixels and a radial exclusion criteria of $\sim$0.7 FWHM. According to our limits, the $L'$-band detection has $J>$13.8 mag\footnote{We acknowledge that this magnitude limit is not yet representative of the expected GPI performances.}.

\subsection{Non-Detection of the $L'$ Counterpart of a Millimeter Emission Source in the Annular Gap}
$H$-band PDI images of HD169142 provided evidence for a low surface brightness annular gap in polarized light which extends from $\sim$40 to 70 AU \citep{Quanz2013b}.
EVLA 7-mm observations revealed an unresolved source (0.10 mJy) in this gap \citep{Osorio2014}, which could be tracing circumplanetary dust
material associated with a forming planet. The angular resolution of the 7-mm map is 0''.23$\times$0''.16 and the signal-to-noise ratio of the compact emission is $\sim$5$\sigma$ above the background. Since no $L'$ counterpart of the mm-emission appears in the $L'$-band images (see Figure~\ref{figure:1}), we can place an upper limit on its luminosity.

To estimate our sensitivity at the expected location of this emission, we inserted a positive artificial planet with increasing flux until we were able to detect it with \texttt{PYNPOINT} and the signal deviated more than 3$\sigma$ from the mean background value.  According to this calculation, the compact source has $L'>$14.0$\pm$0.5 mag. 

\section{Discussion} 
\subsection{The Emission Source in the Inner Cavity}
Given the object's angular separation, the distance to HD169142, and assuming the disk is seen face-on \citep{Quanz2013b}, the physical separation from the central star is 22.7$\pm$4.7 AU. This suggests that it is located within the inner cavity (see Figure~\ref{figure:1}b), right inside the inner edge of the bright rim \citep[$\sim$25 AU,][]{Quanz2013b}. Different scenarios can explain the $L'$-band observations. 

Although unlikely, the detected source could have an instrumental origin. Such a feature could be an AO tip/tilt residual, but it should rotate and be subtracted with the PSF of the star. In case of bad centering behind the AGPM, point-like features can also be generated. However, given the extensive set of tests we did and the data from Biller et al., submitted, we can be confident about the physical origin of the detection.

The bright source could be a background star. However, according to the Besancon galactic model \citep{Robin2003}, in a 1 deg$^2$ portion of the sky at the location of HD169142 we do not expect to have a single object with apparent magnitude $L'\leq$12.2 mag and $J-L'\geq 1.6$ mag.

Due to the limited precision in the astrometry for both the object and the nearby disk structures, we cannot fully exclude that the emission is originated in the disk rim. However, on the one hand, the bright rim is mostly axisymmetric in scattered light (see Figure~\ref{figure:1}b) and there is no maximum in the direction of the \texttt{PYNPOINT} detection. On the other hand, in case of thermal emission, the $L'$-band source must be hotter than $\sim$260 K and smaller than $\sim$1.9 AU, given that it is unresolved in the $L'$-band images.  Such a locally concentrated and warm emission is best explained with a self-luminous compact source.

We can assume that the emission is coming from the photosphere of a companion in quasi-hydrostatic equilibrium undergoing Kelvin-Helmholtz contraction. Under this assumption, the $L'$ luminosity suggests a mass of 35-80 $M_{Jupiter}$ for an age of 3-12 Myrs \citep{Grady2007}, according to the COND models \citep{Baraffe2003}. In this case we should have detected the companion in $J$ band, given our sensitivity at 0''.156 and the predicted $J$-band flux ($J$=13.6-13.7 mag) for such object from the COND models. However, according to the same models, the $J$-band upper limit is still consistent within the uncertainties with the $L'$ photometry and a 28-32 $M_{Jupiter}$ object at 3 Myrs. 
From the theoretical point of view, the presence of a 28-32 $M_{Jupiter}$ companion would be difficult to reconcile with the morphology of the innermost 30 AU.  According to classical gap opening theories \citep[e.g.][]{Lin1993}, we can estimate the width $\Delta$ of the gap if we assume that a single body is carving out the cavity. Under several assumptions, such as the disk scale height at the object's location \citep[2.8 AU from a disk model for HD169142 by][]{Meeus2010}, the geometric factor \citep[$f\approx$0.836,][]{Lin1993}, and the effective disk viscosity ($\alpha\sim$ 0.001), we find a value of $\Delta=$54-59 AU for object masses in the range 28-32 $M_{Jupiter}$. A 10 $M_{Jupiter}$ planet would already be enough to explain the observed cavity size ($\Delta\simeq$25 AU).

Thus, a final possibility could be the detection of a planet during its formation, as it has been proposed for HD100546b \citep{Quanz2013a}. As the star is still accreting, the object itself might still be gathering gaseous material from within the disk cavity. Such an accretion process would increase the observed luminosity due to the presence of a circumplanetary disk, allowing a much lower mass for the object. A drastic increase in the planet luminosity during its formation is expected during the runaway gas accretion phase at a few million years \citep{Marley2007,Mordasini2012}. The $J$-band non-detection would also support the hypothesis of a cooler and/or more extincted object.

\subsection{The Non-Detection in the Annular Gap}
Concerning the non-detection in the $L'$-band images at the location of the 7-mm emission, our dataset allows us to put an upper limit on the mass of a possible object orbiting in the annular gap.  At the distance of the star, our magnitude limit of $L'>$14.0$\pm$0.5 mag would correspond to $<$11-18 $M_{Jupiter}$ at 3-12 Myrs, based on the COND models.

There is therefore a range of possible solutions for the planet-circumplanetary disk system hypothesis. If we assume that the 7-mm flux of 0.10 mJy arises from circumplanetary material emitting like a single-temperature black-body, the non-detection in the $L'$-band implies that it must be cooler than $\sim$240 K and that the emitting area is larger than $\sim$1.5 AU in radius to be consistent with the 7-mm flux. Taking this size as the minimum Hill radius at the expected location of the planet ($\sim$50 AU), we get a lower limit for the planet mass of 0.1 $M_{Jupiter}$.

 In summary, our data are consistent with a planetary mass object with a mass between 0.1-18 $M_{Jupiter}$ surrounded by a circumplanetary disk cooler than $\sim$240 K and with a minimum radius of 1.5 AU. High-spatial resolution (sub-)mm data of the 7-mm source would allow us to test this scenario.

\subsection{Possible multiple planet interaction and evolution}
If the second object is also confirmed, it is interesting to speculate about the possibility of sequential planet formation and how it would affect the evolution of the disk. As suggested by \cite{Bryden2000}, the accumulation of solid particles at the outer edge of a gap, that has been carved out by a protoplanet, could lead to the formation of an additional protoplanetary core at a larger orbital radius. 
\cite{Pierens2008} have shown that when two massive gap-opening planets are embedded in a disk, the gas in between the two gaps is cleared as the two gaps join together. 
The bright rim seen in the scattered light may be the region in between the gaps, before they merge. Such regions consist of gas surface density maxima (and hence pressure maxima) where dust can be trapped. 

Together with HD100546, HD169142 might be a great laboratory to study multiple, and possibly sequential, planet formation empirically.

\section{Conclusions}
In this letter, we present $L'$ and $J$-band observations of HD169142 with the VLT/NACO AGPM coronagraph and GPI, respectively. These images suggest the presence of a low-mass companion in the inner cavity of the transitional disk, at a separation of $\sim$23 AU. Whether this object is a BD or a forming planet still remains to be investigated. In any case, it is likely that this companion affected the disk morphology. 
 Given the proper motion of HD169142, observations as early as mid-2015 will allow us to rule out the hypothesis of a background source. Upcoming instruments, such as VLT/SPHERE, will be crucial for proper motion confirmation and follow-up. Furthermore, given the expected orbital time of the object ($\sim$86 years), it should move of $\sim 40^{\circ}$ in 10 years.

Finally, our images do not exclude the possibility of a second object (0.1-18 $M_{Jupiter}$) forming in the annular gap (40-70 AU), as suggested by recent millimeter observations \citep{Osorio2014}. If future observations (e.g. mm and sub-mm data) confirm it, HD169142 would be forming a planetary system with at least two planets, and would boost our understanding of multiple and possibly sequential planet formation.

\acknowledgments
Based on observations obtained at the Gemini Observatory, operated by the Association of Universities for Research in Astronomy, Inc., under a cooperative agreement with the NSF on behalf of the Gemini partnership: NSF (United States), the National Research Council (Canada), CONICYT (Chile), the Australian Research Council (Australia), Minist\'erio da Ci\^encia, Tecnologia e Inova\c{c}\~{a}o (Brazil) and Ministerio de Ciencia, Tecnolog\'ia e Innovaci\'on Productiva (Argentina). 
We thank Bruce Macintosh and the GPI team for support during the observations.
GA, CC-G, MO, and JMT acknowledge support from MINECO (Spain) AYA2011-30228-C03 grant.

{\it Facilities:} \facility{VLT:Yepun (NACO)}.

\clearpage

\end{document}